\documentclass[aps,prl,showpacs,twocolumn,groupedaddress]{revtex4}  
\usepackage{graphicx}  
\usepackage{dcolumn}   
\usepackage{bm}        
\usepackage{amssymb}   

\begin{document}


\hspace{5.2in}

\title{\boldmath  
Probing $WW\gamma \gamma$ and $ZZ\gamma \gamma$ 
quartic anomalous couplings with 10 pb$^{-1}$ at the LHC  \\}
\author{E. Chapon}\email{emilien.chapon@cea.fr}
\affiliation{CEA/IRFU/Service de physique des particules, CEA/Saclay,  91191 Gif-sur-Yvette cedex, France}
\author{O. Kepka}\email{kepkao@fzu.cz}
\affiliation{CEA/IRFU/Service de physique des particules, CEA/Saclay,  91191 Gif-sur-Yvette cedex, France}
\affiliation{IPNP, Faculty of Mathematics and Physics,
Charles University, Prague} 
\affiliation{Center for Particle Physics, Institute of Physics, Academy of Science, Prague} 
\author{C. Royon}\email{royon@hep.saclay.cea.fr}
\affiliation{CEA/IRFU/Service de physique des particules, CEA/Saclay,  91191 Gif-sur-Yvette cedex, France}
\begin{abstract}
We report on a possible measurement at the LHC using the first data and a
luminosity of 10 pb$^{-1}$ of $W$ and $Z$ pair production via two-photon
exchange. This measurement
allows to increase the present sensitivity on $WW\gamma \gamma$ and
$ZZ\gamma \gamma$ quartic anomalous couplings 
from the LEP experiments by
almost three orders of magnitude.
\end{abstract}

\maketitle

In the Standard Model (SM) of particle physics, the couplings of fermions and 
gauge bosons are constrained by the gauge symmetries of the Lagrangian.
The measurement of $W$ and $Z$ boson pair productions via the exchange of
two photons  
allows to provide directly stringent tests
of one of the most important and least understood
mechanism in particle physics, namely the
electroweak symmetry breaking~\cite{stirling}. The non-abelian gauge nature of the SM
predicts the existence of quartic couplings
$WW\gamma \gamma$
between the $W$ bosons and the photons which can be probed directly at the 
Large Hadron Collider (LHC) at CERN.
The quartic coupling to the $Z$ boson $ZZ\gamma \gamma$ is not present in the
SM. 

The quartic couplings test 
more generally new physics which couples to electroweak bosons.
Exchange of heavy particles beyond the SM might manifest itself as a
modification of the quartic couplings appearing in contact 
interactions~\cite{higgsless}. It is
also worth noticing that in the limit of infinite Higgs masses, or in Higgs-less
models~\cite{higgsless}, new structures not present in the tree level Lagrangian
appear in the quartic $W$ coupling. For
example, if the electroweak breaking mechanism does not manifest itself in the
discovery of the Higgs boson at the LHC or supersymmetry, the presence of
anomalous --- non SM like --- couplings might be the first evidence of new
physics in the electroweak sector of the SM.

In this Letter, we will demonstrate that it is possible to probe quartic
couplings between the $W$ or $Z$ and the photons at the LHC using the first data
at low luminosity. The LHC is a proton proton machine with a nominal
center-of-mass energy of 14 TeV, located at CERN, Geneva, Switzerland. The first
collisions are expected to occur towards the end of 2009 at a reduced center-of-mass
energy of 10 TeV. A typical luminosity of 10 (100) pb$^{-1}$ is
expected to be accumulated in a few days (weeks). 

\begin{figure}
\includegraphics[scale=0.6]{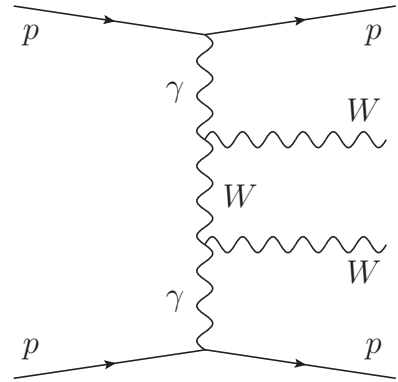}
\caption{\label{fig1} 
Diboson production through two photon exchange. The intact protons
in the final state are scattered at very small angles. 
}
\end{figure}

High energy colliders such as the incoming LHC are the natural place to look
for anomalous quartic couplings between the photon and the $W$ or $Z$ bosons.
The process we want to study at the LHC is depicted in Fig.~\ref{fig1}, and
corresponds to $pp \rightarrow pWWp$. In this photon-induced process, the two
quasi-real 
photons interact through the exchange of a virtual $W$, leading to a pair of $W$s 
in the final state. The advantage of this kind of events is that they are
extremly clean, there are two $W$s (or $Z$s) which can be detected in the ATLAS or CMS
central detectors while the intact proton leave undetected in the beam
pipe. The cross section of this Quantum Electrodynamics (QED) process in the
standard model is known precisely and is equal to 62 fb at a center-of-mass
energy of 10 TeV.

In this Letter, we restrict ourselves to the implementation of the genuine
quartic anomalous 
$\gamma \gamma WW$ and $\gamma \gamma ZZ$ using the lowest dimension
operators possible in the Lagrangian. They are of dimension six and preserve
by construction the
custodial $SU(2)_c$ symmetry required to keep the $\rho=
M_W^2/(M_Z^2 ~\cos^2 \theta_W)$ parameter to its
experimental value close to 1 where $M_W$, $M_Z$ and $\theta_W$ are respectively the 
$W$ and $Z$ boson masses, and the weak mixing angle. In addition, they are
not related to the anomalous triple gauge couplings in any way.
The most general Lagrangians leading to anomalous $WW\gamma \gamma$ and
$ZZ \gamma \gamma$ quartic couplings are
the following~\cite{anomalous}
\begin{eqnarray}
     \mathcal{L}_6^0 &=& \frac{-e^2}{8} \frac{a_0^W}{\Lambda^2} F_{\mu\nu} F^{\mu\nu} W^{+\alpha} 
     W^-_\alpha \nonumber \\
     &~&- \frac{e^2}{16\cos^2 \theta_W} \frac{a_0^Z}{\Lambda^2} F_{\mu\nu} F^{\mu\nu} Z^\alpha
Z_\alpha\nonumber \\
     \mathcal{L}_6^C & = & \frac{-e^2}{16} \frac{A_C^W}{\Lambda^2} F_{\mu\alpha} F^{\mu\beta} 
     (W^{+\alpha} W^-_\beta + W^{-\alpha} W^+_\beta) \nonumber \\
&~&	- \frac{e^2}{16\cos^2 \theta_W} \frac{a_C^Z}{\Lambda^2} F_{\mu\alpha} F^{\mu\beta} Z^\alpha Z_\beta\nonumber\\
\label{eq:anom:lagrqgc}
\end{eqnarray}
where $W^{\pm \alpha}$ and $Z^{\alpha}$ are the $W$ and $Z$ boson fields, and 
$F_{\mu \nu}= \partial_{\mu} A_{\nu} - \partial_{\nu} A_{\mu}$ the
electromagnetic tensor. The new scale $\Lambda$ is
introduced so that the Lagrangian has the correct dimension 4 and can be
interpreted as the scale of new physics.
In this Lagrangian, the $W$ and
$Z$ parts are allowed to have their specific parameters $a_0^W$, $a_C^W$, $a_0^Z$
and $a_C^Z$. Such Lagrangian conserves C, P and T parities separately, and
represents the most natural extension of the SM.

The current best 95\% confidence level (C.L.) limits on the parameters of quartic anomalous couplings were
determined by the OPAL Collaboration~\cite{opal} where the quartic couplings 
were measured in $e^+e^-\rightarrow W^+W^-\gamma$, $e^+e^-\rightarrow \nu\bar{\nu}\gamma\gamma$ (for $WW\gamma\gamma$ anomalous
couplings), and $e^+e^-\rightarrow q\bar{q} \gamma\gamma$ (for $ZZ\gamma\gamma$ couplings) 
with center-of-mass energies up to 209 GeV and are given in Table~\ref{table2}.

The scattering amplitudes are quadratically divergent for all anomalous coupling
parameters. The steep rise of the cross section at high energy leads 
immediately to the violation of unitarity. The tree level unitarity uniquely 
restricts the $WW\gamma\gamma$ couplings to the SM values at asymptotically 
high energies. This implies that any deviation of 
the anomalous parameters $a_0^Z/\Lambda^2$, $a_C^Z/\Lambda^2$, 
$a_0^W/\Lambda^2$, $a_C^W/\Lambda^2$ from the SM zero value has to be 
multiplied by a form factor which vanishes in the high energy limit and which 
introduces a regularization of the cross section at the new physics scale. 
At LEP, where the center-of-mass energy was rather low, the wrong high energy 
behaviour did not violate unitarity; however, it must be reconsidered at the 
LHC. We therefore modify the couplings  
using the form factors that leave the couplings at small energies the 
same but suppress their effect when the center-of-mass energy $W_{\gamma\gamma}$ 
increases~\cite{piotr}, such as
\begin{eqnarray}
a \rightarrow \frac{a}{(1+W^2_{\gamma\gamma}/\Lambda^2)^n}.
\label{eq:anom:formfactor}
\end{eqnarray}
The exact form of the form factor is not imposed but rather only 
conventional and the same holds for a value of the exponent $n$.
$\Lambda^2$ corresponds to the scale where new physics should appear.
At the LHC, usual values for the $n$ and $\Lambda$ parameters are respectively 2
and 2 TeV~\cite{piotr}. The cross sections corresponding to the Lagrangian 
of (\ref{eq:anom:lagrqgc}) for $WW$ and $ZZ$ productions 
as well as the different backgrounds
were implemented in the Forward Physics Monte Carlo (FPMC)~\cite{fpmc}. Technically,
the scattering amplitudes were computed using the CompHEP~\cite{comphep} Monte
Carlo and implemented in FPMC so that the decay and hadronization of the
different particles can be performed using HERWIG~\cite{herwig}.

\begin{figure}
\includegraphics[scale=0.42]{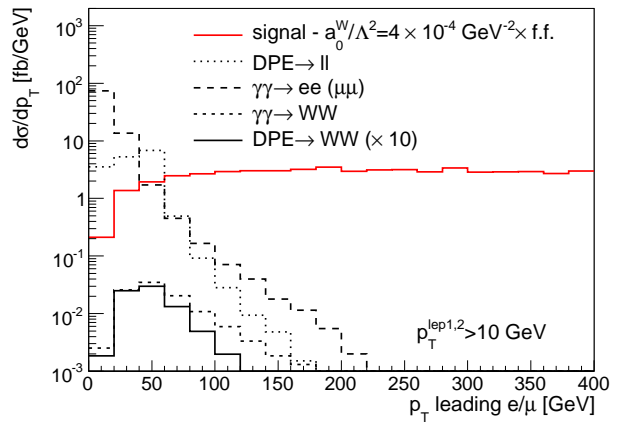}
\caption{\label{fig2} Transvserse momentum of the leading lepton
for the quartic anomalous coupling signal and the different SM backgrounds.
}
\end{figure}

In the following, we will study the sensitivity on quartic anomalous 
$a_0^W$ and $a_C^W$ couplings
using the first data to be taken at the LHC, and namely a luminosity of 10 (or
100) pb$^{-1}$ which can be accumulated in a couple of days or weeks at a
center-of-mass energy of 10 TeV. The results were obtained using a fast
simulation of the ATLAS detector using ATLFAST$++$~\cite{atlfast} and similar results could be
obtained using the CMS detector. We are considering $W$ pair production in the
ATLAS detector (see Fig.~\ref{fig1}), and to simplify the study, we limit
ourselves to the leptonic decays of the $W$ pair to electrons or muons. This
avoids many issues to be considered in the first days of data taking in the
ATLAS or CMS experiments concerning the jet energy calibration or the jet
triggers when one uses the hadronic decays of the $W$ pairs.
The signal is characterised by the presence of two high transverse
momentum ($p_T$) leptons
(electrons or muons) reconstructed in the ATLAS central detector, and the 
absence of any other reconstructed object or energy flow since the scattered
protons leave undetected in the beam pipe. The signature of such events is
thus very clear and the potential inclusive and exclusive backgrounds can be easily rejected
as we will see in the following. To trigger on such events is also simple since a
high $p_T$ single lepton or dilepton unprescaled trigger can be used.

Different backgrounds have to be taken into account for the $WW$ production,
namely the non-exclusive $W$ pair production, the dilepton production through
$\gamma$ or Pomeron exchange, and the $W$ diffractive production. 
The $p_T$
distribution of the events with a $WW\gamma \gamma$ anomalous coupling
$a_0^W=4$ 10$^{-4}$ GeV$^2$ (which means a value about two orders of
magnitude smaller than the LEP limit) is given in Fig.~\ref{fig2}, together
with the different backgrounds. After requesting the presence of two 
reconstructed leptons (electrons or muons) in the main ATLAS detector of 
$p_T>10$ GeV, the number of background events for 10 pb$^{-1}$ is respectively
17.4, 6.0, 0.003 and 0.03 for dilepton production through $\gamma$ exchange,
dilepton diffractive production, and $W$ pair production via Pomeron or 
photon exchange in the detector acceptance.
In Fig~\ref{fig2}, we already see that the $p_T$ of the leading
lepton extends to much higher values for the signal events than for the
background. 

\begin{figure*}
\includegraphics[scale=0.42]{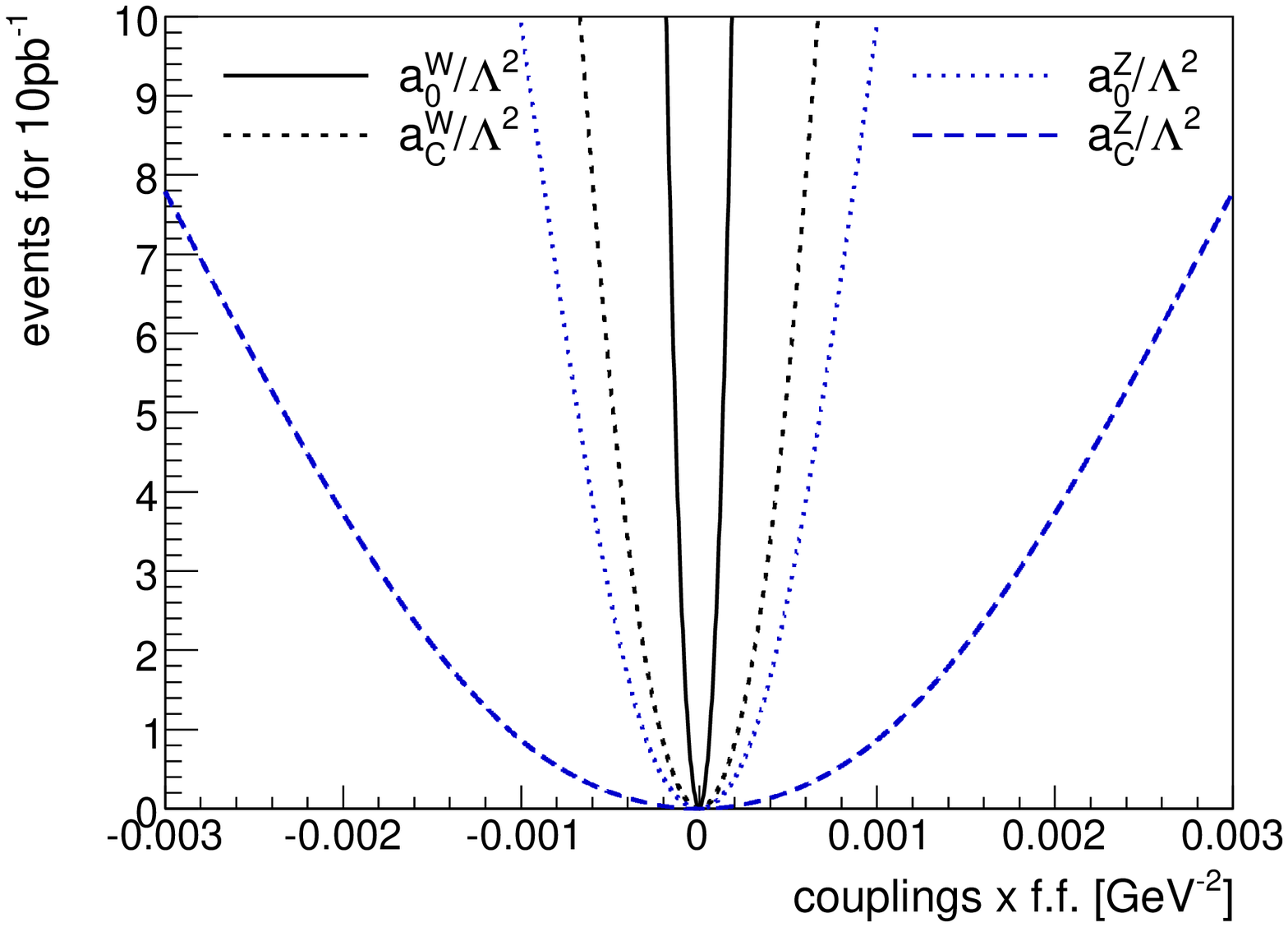}
\includegraphics[scale=0.42]{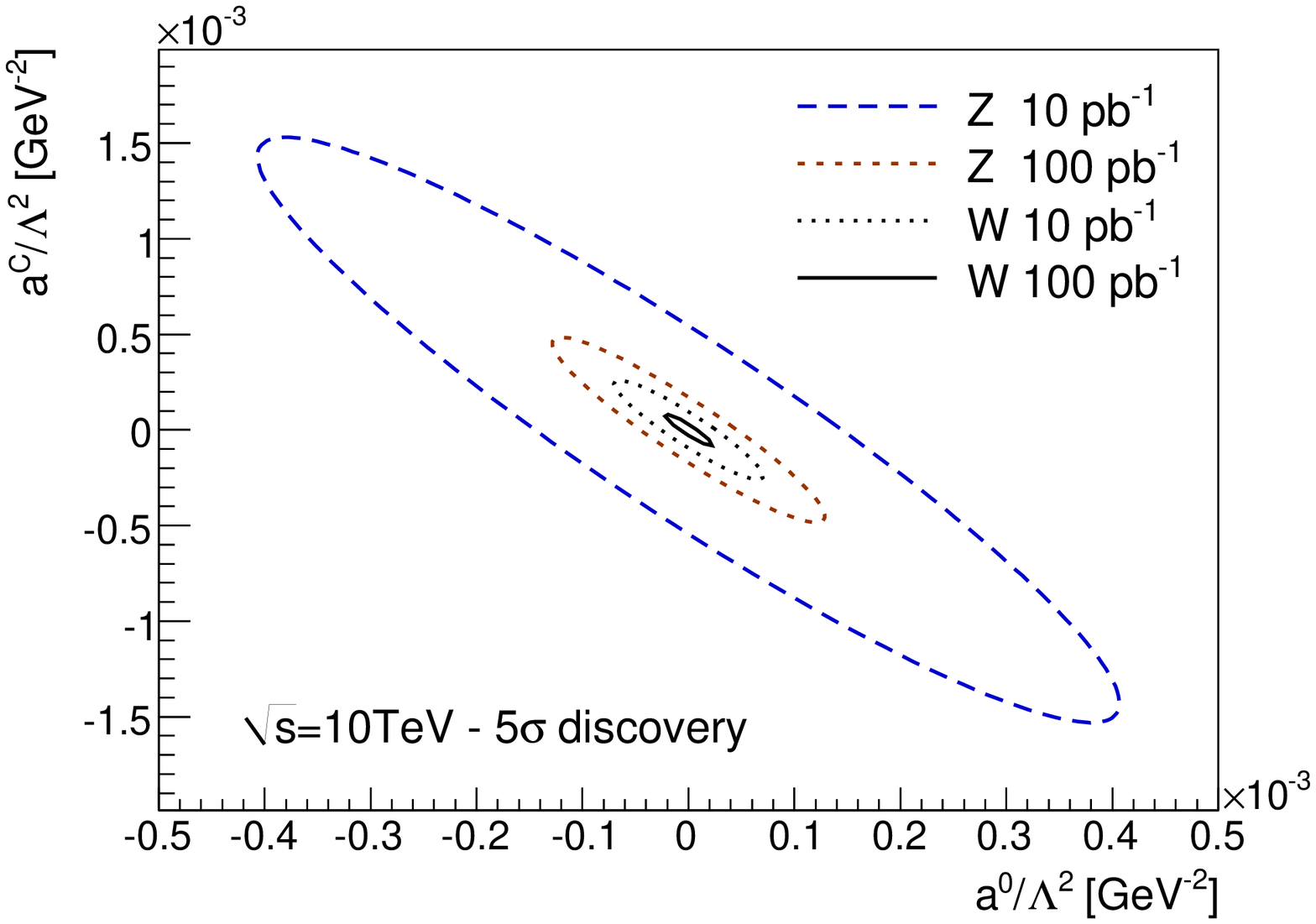}
\caption{\label{fig3} Left: Number of signal events after cuts as a function of the 
value of the quartic anomalous coupling for 10 pb$^{-1}$ at the LHC.
Right: $5\sigma$ discovery contours for the $WW$ and $ZZ$ quartic 
anomalous couplings at $\sqrt{s}=10$ TeV for luminosities of 10 and 
100 pb$^{-1}$.}
\end{figure*}

Let us now discuss each background in turn. The non-diffractive
$W$ pair production is suppressed by requesting the presence of two leptons and
nothing else in the ATLAS detector since the inclusive background always shows
some hadronic activity in the calorimeter or in the forward part of
the tracking detector. This background is thus found to be negligible after the
exclusivity cut. It is worth noticing that this cut requires that we consider only
low instantaneous luminosity when only one interaction occurs per bunch
crossing. At higher instantaneous luminosity, the proton tagging using dedicated
detectors will be needed~\cite{us}. The dilepton production through photon 
exchange (QED process) is
suppressed by requesting the presence of a leading lepton with $p_T>160$ GeV,
and of missing energy greater than 20 GeV, which is natural when one requests the
presence of two $W$s. The pure SM $W$ pair background (without any anomalous
couplings) via photon exchange is small
since the value of the cross section
is low (62 fb) and it is further suppressed by requiring a reconstructed lepton
with $p_T>160$ GeV.
The diffractive production of dileptons or $W$ pairs via double pomeron
exchange (DPE) were
studied using the FPMC Monte Carlo. The quark and gluon structure of the Pomeron
was taken using the H1 parametrisation of the Pomeron~\cite{h1zeus}
with a survival probability at the LHC of 0.03~\cite{survival} (the survival
probabilty in the case of $\gamma$-induced processes is 0.9). This background
suffers more uncertainties than the photon exchange processes since it is less
understood theoretically but it leads to a negligible background after the exclusivity
cut. After all cuts, the background is found to be negligible. As an example, the 
number of events for signal
for $a_0^W/\Lambda^2$=2. 10$^{-4}$ is 19 before cuts in the detector acceptance and 12 
after cuts, showing that the signal
is not much affected by the cuts we introduced.
The number of events after the different cuts as a function of the
value of the anomalous coupling is given in Fig.~\ref{fig3}, left, for a
luminosity of 10 pb$^{-1}$. We already see that the reach on the anomalous
coupling $a_0^W$ and $a_C^W$ is expected to be much better than at LEP by more
than two orders of magnitude.

We also studied the effect of the anomalous $ZZ \gamma \gamma$ couplings $a_0^Z$
and $a_C^Z$. The cuts are even simpler since two leptons of the same charge are
produced when the $Z$ bosons decay leptonically. We thus request two like sign
leptons or three reconstructed leptons and the $p_T$ of the leading lepton is
larger than 100 GeV. All background is found to be negligible after those cuts.

The sensitivity on anomalous coupling is depicted in Fig.~\ref{fig3}, right,
where the 5$\sigma$ discovery contours are displayed in the plane ($a_0^W$,
$a_C^W$), or ($a_0^Z$, $a_C^Z$) for two different values of the accumulated
luminosity of 10 and 100 pb$^{-1}$. The reach on the different anomalous
couplings is also given in Table~\ref{table2}. At LHC energies,
the cross sections are the same for negative and positive values of the
anomalous couplings and the obtained sensitivities are symmetric.
We note that the LEP sensitivity on quartic
anomalous couplings can be increased by two or three orders of magnitude
depending on the coupling with a low luminosity of 10 or 100 pb$^{-1}$.
By comparison, the gain on trilinear gauge coupling is not that stringent with
respect to LEP~\cite{olda}.

\begin{table}[htb]
\centerline{
   \begin{tabular}{|c||c|c|c|}
    \hline
    \raisebox{-1.5ex}[0pt][0pt]{Couplings} & 
    OPAL limits & 
    \multicolumn{2}{c|}{Sensitivity @ $\mathcal{L} = 10$ (100) pb$^{-1}$} \\
    &  \small[GeV$^{-2}$] & 5$\sigma$ & 95\% CL \\ 
    \hline
    $a_0^W/\Lambda^2$ & [-0.020, 0.020] & 2.2 10$^{-4}$ & 1.0 10$^{-4}$\\
                      &                 & (7.3 10$^{-5}$) & (3.3 10$^{-5}$)\\ \hline               
    $a_C^W/\Lambda^2$ & [-0.052, 0.037] & 5.9 10$^{-4}$ & 3.5 10$^{-4}$\\
                      &                 & (2.4 10$^{-4}$) & (1.1 10$^{-4}$)\\ \hline               
    $a_0^Z/\Lambda^2$ & [-0.007, 0.023] & 1.0 10$^{-3}$ & 5.2 10$^{-4}$\\
                      &                 & (3.7 10$^{-4}$) & (1.7 10$^{-4}$)\\ \hline               
    $a_C^Z/\Lambda^2$ & [-0.029, 0.029] & 3.0 10$^{-3}$ & 1.8 10$^{-3}$\\
                      &                 & (1.3 10$^{-3}$) & (5.9 10$^{-4}$)\\ \hline               
    \hline
   \end{tabular}
   }
\caption{Limits on anomalous coupling coming
from the LEP OPAL experiment. 
Sensitivity at low luminosity. The $5\sigma$ discovery potentials as well
as the 95\% C.L. limits are given for 10 pb$^{-1}$ and 100 pb$^{-1}$ in
parenthesis.}
\label{table2}
\end{table}

In this Letter, we described a new study to be performed using the first data at
the LHC at a center-of-mass energy of 10 TeV using a luminosity of 10 or 100
pb$^{-1}$. Through the measurements of $WW$ and $ZZ$ 
productions via photon exchange at the LHC when
the $Z$ or $W$ bosons decay leptonically, it is possible to improve the
sensitivity on the $\gamma \gamma WW$ and $\gamma \gamma ZZ$ anomalous couplings
by almost three orders of magnitude with respect to the LEP sensitivity. This
will be one of the most stringent tests of the electroweak symmetry breaking at
the beginning of the LHC which will happen before the search for the Higgs boson
and might lead to a first evidence for effects beyond the Standard Model via the
indirect effects of a heavy scalar particle or an indirect test
of Higgs-less models.

\section*{Acknowledments}
We thank K. Piotrzkowski and T. Pierzchala for useful discussion about the
CompHep program.

\end{document}